\documentclass[twocolumn,amsmath,amssymb,10pt,a4paper]{revtex4}
\pdfoutput=1

\usepackage{hyperref}

\hypersetup{
        colorlinks=true,
        linktoc=all,
        linkcolor=violet,
        urlcolor=blue
}

\usepackage{geometry}
\usepackage{amsmath, amssymb}
\usepackage{graphicx}
\usepackage{bbold}
\usepackage{algorithm}
\usepackage{algpseudocode}
\usepackage{amsthm}
\usepackage{mathtools}
\usepackage{MnSymbol}

\theoremstyle{definition}
 
\theoremstyle{remark}

\newcommand{\be}{\begin{equation}}
\newcommand{\ee}{\end{equation}}
\newcommand{\bd}{\begin{displaymath}}
\newcommand{\ed}{\end{displaymath}}
\newcommand{\BE}{\begin{eqnarray}}
\newcommand{\EE}{\end{eqnarray}}

\newcommand{\mcN}{\mathcal{N}}
\newcommand{\mcP}{\mathcal{P}}
\newcommand{\mcR}{\mathcal{R}}

\graphicspath{{../../figs/}}

\usepackage{epigraph,varwidth}
\usepackage{etoolbox}
\usepackage{lipsum}

\usepackage{tcolorbox}
\newtcolorbox{mybox}{colback=green!5!white,colframe=green!75!black}

\begin{document}

\title{``Negative probabilities'' as relative probabilistic expressions} 

\author{John Realpe-G\'omez$^1$}\email{john.realpe@gmail.com} 
\affiliation{Instituto de Matem\'aticas Aplicadas, Universidad de Cartagena, Bol\'ivar 130001, Colombia}

\date{\today}

\begin{abstract}
Here we briefly discuss how negative numbers, or ``negative probabilities'', can naturally arise in probabilistic expressions and be given an operational interpretation. Like the use of negative numbers in arithmetical expressions, the use of negative probabilities can have substantial practical value. Indeed, some of the ideas discussed here have led to stochastic simulation algorithms that require much less memory than the best classical algorithms known to date [\href{https://arxiv.org/abs/1906.00263}{\em arXiv:1906.00263}]. 
\end{abstract}
\maketitle

The idea of negative numbers in probabilistic expressions, or ``negative probabilities'', arose in quantum physics~\cite{wigner1932quantum,dirac1942physical,feynman1987negative}. More recently, some authors~\cite{burgin2010interpretations,abramsky2014operational} have explored the formalization of negative probabilities in a classical context by adding a bit of information, $\tau = \pm 1$, that essentially attach a type to each sample generated, qualifying it as an actual sample ($\tau = +1$) or an anti-sample ($\tau = -1$). Anti-samples can be represented, for instance, by the same type of object representing the stochastic variable under consideration, e.g. a coin, but with a different color. 
The key point is that samples and anti-samples in the same state anihilate each other, i.e. they are both removed from the ensemble simultaneously.

Here we briefly discuss how negative probabilities could be understood as probabilistic expressions relative to another reference probabilistic expression. Moreover, we argue that it is not always necessary to work with a duplicated sample space as in Refs.~\cite{burgin2010interpretations,abramsky2014operational}. Rather, negative probabilities can in some cases be interpreted as corrections that need to be applied to a reference ensemble to transform it into a target ensemble. Or, equivalently, as corrections that need to be applied to a single sample, generated according to a reference probability distribution, to transform it into a sample generated according to a target probability distribution.

Unnormalized probabilistic expressions could also be given a simple operational interpretation as referring to intermediate ensembles with a number of samples different from that of a target ensemble. Such a different number of samples can be associated to the replication or removal of samples. At some point we have to add or remove further samples (equivalently, add anti-samples) to recover the actual number of samples of the target ensemble. 

Aiming at practical applications, here we try to discuss these ideas in a hopefully intuitive way, using examples as simple as possible, without worrying about the abstract formalism underneath. We do not claim any mathematical rigor. Indeed, in some of the sampling methods we discuss there may be errors that happen with a small probability as long as the number of samples generated is large enough. Our main purpose is to bring awareness of the potential benefits of working with negative probabilities. We leave a more rigorous presentation for future work. 

Essentially, signed and unnormalized probability distributions allow for an operational interpretation of linear algebra manipulations on probability distributions, without imposing the usual constrains that all the numbers involved should be in the interval $[0,1]$ nor that all expressions need be normalized.  Feynman~\cite{feynman1987negative} had some early intuition on this possibility, building on a comparison to the standard use of negative numbers in arithmetical expressions. Indeed, negative numbers in arithmetical expressions remind us of our ``deterministic'' debts, i.e., fixed debts without stochastic variations. Similarly, negative probabilities can remind us of our ``statistical debts''. 

Let us first discuss the case where signed probability distributions can be considered as relative to a reference probability distribution. Consider a biased coin which, when tossed, lands heads or tails with probability $1-p$ and $p$, respectively. So, the coin is described by the probability vector
\be\label{e:coin_prob}
\mcP_{\rm coin} = \begin{pmatrix} 1-p \\ p\end{pmatrix} = \mathcal{R}_{\rm coin} + \mathcal{N}_{\rm coin}, 
\ee
where 
\BE
\mathcal{R}_{\rm coin} &=& \frac{1}{2}\begin{pmatrix} 1 \\ 1\end{pmatrix} , \\
\mathcal{N}_{\rm coin} &=&  \frac{1-2 p}{2}\begin{pmatrix} 1 \\ -1\end{pmatrix},
\EE
and the first and second entries of the vectors involved refer to heads and tails, respectively. The right hand side can be formally understood as an abstract change of basis. We have introduced this change of basis to show that it can also be interpreted in probabilistic terms, even though there are negative numbers involved.

Indeed if we multiply Eq.~\eqref{e:coin_prob} by a large natural number $M$, we can interpret each of its terms as an ensemble of $M$ coins. The vector $M \mcP_{\rm coin}$  would describe the original ensemble of $M$ biased coins. We refer to this as the target ensemble. The vector $M \mathcal{R}_{\rm coin}$ would describe a flat ensemble of $M$ unbiased coins, i.e., distributed uniformly at random. We would refer to this as the reference ensemble. The vector $M \mathcal{N}_{\rm coin}$, which has negative numbers, can be interpreted~\cite{realpe2019quantum} as corrections that should be applied to the flat reference ensemble of unbiased coins, $M \mathcal{R}_{\rm eff}$, to recover the original ensemble of biased coins, $M \mcP_{\rm coin}$. The negative numbers can be interpreted as referring to anti-samples instead of samples, as done in Refs.~\cite{burgin2010interpretations,abramsky2014operational}. Alternatively, we can give an interpretation as transitions that samples from the reference ensemble undergo from states associated to negative entries in $\mathcal{N}_{\rm coin}$, here tails, into states associated with positive entries in $\mathcal{N}_{\rm coin}$, here heads. A detailed example of this alternative approach can be found in Ref.~\cite{realpe2019quantum}.

For concreteness, let us assume that $1-2p>0$. To generate an ensemble of $M$ biased coins, described by $M\mathcal{P}_{\rm coin}$, we can first generate an ensemble of $M$ unbiased coins, described by $M\mathcal{R}_{\rm coin}$. Afterwards we need to generate on average $M (1-2p)/2$ new coins in state heads and remove the same number of coins in state tails, so the ensemble remains with $M$ coins---this is manifested in that the sum of the entries of the vector $\mathcal{N}_{\rm coin}$ is zero. We can implement this process via transitions that samples undergo from states associated to negative entries in $\mathcal{N}_{\rm coin}$, here tails, into states associated with positive entries in $\mathcal{N}_{\rm coin}$, here heads. More precisely, by fliping coins in state tails with probability $1-2p$, we can transform $M\mathcal{R}_{\rm coin}$ into $M \mcP_{\rm coin}$~\cite{realpe2019quantum} (cf. Fig.~4 and Sec.~III~B~1 therein). Since, on average half of the coins in the reference ensemble, $M\mcR_{\rm coin}$, are in state tails, this process flips on average $M(1-2p)/2$ coins as expected.

We can interpret signed probability distributions like $\mathcal{N}_{\rm coin}= \mcP_{\rm coin} - \mathcal{R}_{\rm coin}$ as probabilistic expressions encoding the stochastic behavior of the original system, described by $\mcP_{\rm coin}$, relative to a reference probability distribution,  $\mathcal{R}_{\rm coin}$. In other words, the signed probability distribution $\mathcal{N}_{\rm coin}$ describes a stochastic dynamics of samples that are added to and removed from the reference ensemble $M\mcR_{\rm coin}$. In this sense, $M\mcR_{\rm coin}$ could be considered analogous to the Dirac sea~\cite{dirac1930theory} in relativistic quantum mechanics, which allows anti-particles to be interpreted as holes in it. If $M$ is large enough, the probability to remove all samples from the reference ensemble is very small. Alternatively, such dynamics can be described as transitions of the samples in $M\mcR_{\rm coin}$ from a given state, associated to negative numbers in $\mcN_{\rm coin}$ to another state, associated with positive numbers in $\mcN_{\rm coin}$ as we discussed above (see also Ref.~\cite{realpe2019quantum}). In this case the number of samples remains constant. 

These ideas can in principle be generalized to any probability distribution, $\mathcal{P}$, and any reference probability distribution, $\mathcal{R}$ or conditional probability distributions. In the latter case we apply the same tools to each probability distribution that arises from fixing the conditioned variable to a given value~\cite{realpe2019quantum} (see Sec.~III~C~2 and Appendix~A therein).

Let us now discuss the case where the signed probability distribution can be considered as relative to a reference unnormalized probability distribution. Equation~\eqref{e:coin_prob} can also be written as 
\be\label{e:coin_prob'}
{\mcP}_{\rm coin} = \widetilde{\mathcal{R}}_{\rm coin} + \widetilde{\mathcal{N}}_{\rm coin}, 
\ee
where
\BE
\widetilde{\mathcal{R}}_{\rm coin} &=& \frac{1-f}{7}\begin{pmatrix} 4 \\ 3\end{pmatrix},\\
\widetilde{\mathcal{N}}_{\rm coin} &=& f\begin{pmatrix} -3 \\ 4\end{pmatrix},
\EE
and $f= (7 p - 3)/25$. Equation~\eqref{e:coin_prob'} can also be formally interpreted as an abstract change of basis different from that in Eq.~\eqref{e:coin_prob}. We introduce this new change of basis to show that it can also be interpreted probabilistically, even though the reference expression, $\widetilde{\mathcal{R}}_{\rm coin}$, is not normalized in this case, unless $f=0$. 

Indeed, although the vector $\widetilde{\mathcal{R}}_{\rm coin}$ contains only positive entries, their sum is $1- f = 1+(3-7 p)/25$, which is equal to one only when $p=3/7$. However, if we multiply Eq.~\eqref{e:coin_prob'} by a large natural number $M$, we can still interpret $\widetilde{\mathcal{R}}_{\rm coin}$ as a reference ensemble of $(1-f) M$ coins distributed according to probability vector $({4}/{7}, 3/7)$. The entries of the vector $\widetilde{\mathcal{N}}_{\rm coin}$ add up to $f$, so it compensates for the different number of samples in the reference ensemble, $M \widetilde{\mathcal{R}}_{\rm coin}$, by adding (if $f>0$) or removing (if $f<0$) a number $|f M|$ of coins.

For concreteness, let us assume $p<3/7$, so $f <0$. To generate an ensemble with an average number of $M$ coins distributed according to $\mcP_{\rm coin}$ we can first generate a reference ensemble with $(1-f) M > M$ coins, distributed according to the probability vector $\widetilde{\mathcal{R}}_{\rm coin}/(1-f) = (4/7 , 3/7)$. The vector $\widetilde{\mathcal{N}}_{\rm coin}$ can be interpreted as removing on average $-4 f M >0$ coins in state tails (assuming there are enough), associated to a negative entry of $\widetilde{\mathcal{N}}_{\rm coin}$, and adding on average $-3 f M>0$ coins in state heads, associated to a postivie entry of $\widetilde{\mathcal{N}}_{\rm coin}$. Besides recovering the original bias of the probability vector $\mcP_{\rm coin}$, this would remove an average number of $-f M$ coins from the reference ensemble, $M \widetilde{\mathcal{R}}_{\rm coin}$, leaving on average only $M$ coins as expected.  In case, of an extremely large deviation of the ensemble $M\widetilde{\mcR}_{\rm coin}$ such that there are no enough coins in state tails to remove, we can remove those available and afterwards remove coins at random until we get down to $M$ coins. This is a rare event with a very small probability as long as $M$ is large enough.

Although our focus has been on the ensemble level, these ideas can also be applied iteratively, sample by sample. An example is the general memory-enhanced stochastic algorithm recently introduced in Ref.~\cite{realpe2019quantum} (see Fig.5, Sec.~III~C~2, and Appendix A therein). This example deals with the case where the target and the reference probabilistic expressions are conditional probability distributions. The case where the reference expression is an unnormalized probability distirbution requires that we introduce a probability of replication and removal of samples. This makes the use of the sample/anti-sample formalism~\cite{burgin2010interpretations,abramsky2014operational} more convenient in this case, at the expense of duplicating the sample space.

For instance, to generate a sample from $\mcP_{\rm coin}$ based on Eq.~\eqref{e:coin_prob'} we can first generate a sample according to the probability vector $\widetilde{\mathcal{R}}_{\rm coin}/(1-f) = (4/7 , 3/7)$.  Afterwards, with probability $|f|<1$ we can replicate the sample (if $f<0$) or remove it (if $f>0$). Were we to repeat this process $M$ times we would obtain an ensemble of $(1-f)M$ samples, on average, distributed according to $\widetilde{\mathcal{R}}_{\rm coin}/(1-f)$. However, this is not our aim as we still need to take into account the vector $\widetilde{\mathcal{N}}_{\rm coin}$. So, with probability $|3f|< 1$ we add (if $f<0$) or remove (if $f>0$) a sample in state heads. Similarly, with probability $|4f|<1$ we remove (if $f<0$) or add (if $f>0$) a sample in state tails. If at some point in this process there are no samples to remove, we create an anti-sample in the corresponding state This anti-sample will anihilate along with a new sample in the same state as soon as the latter appears. In principle, we can also deal with this case without the need of anti-samples along the lines of Ref.~\cite{realpe2019quantum}. It is not clear, though, this would be the case when $|f|>1$, where we would need to replicate or remove samples, or add anti-samples, more than once at each step. 

More generally, given a generic probability distribution $\mcP$, we can define a signed probability distribution $\widetilde{\mathcal{N}} = \mcP -\widetilde{\mathcal{R}} $ relative to a non-negative reference vector or function $\widetilde{\mathcal{R}}$. This reference function may or may not be a probability distribution depending on whether it is normalized or not. 

Finally, let us discuss the more general case where the signed probability distribution need not necessarily be relative to a reference probabilistic expression. Equation~\eqref{e:coin_prob} can also be written as 
\be\label{e:coin_prob''}
{\mcP}_{\rm coin} = \widetilde{\mathcal{N}}_{\rm coin}^{(1)} + \widetilde{\mathcal{N}}_{\rm coin}^{(2)}, 
\ee
where
\BE
\widetilde{\mathcal{N}}_{\rm coin}^{(1)} &=& \begin{pmatrix} 2 \\ -p\end{pmatrix},\label{e:Ncoin1}\\
\widetilde{\mathcal{N}}_{\rm coin}^{(2)} &=&  \begin{pmatrix} -1- p \\ 2p \end{pmatrix}.\label{e:Ncoin2}
\EE
Here there is no non-negative expression that like $\mathcal{R}_{\rm coin}$ in Eq.~\eqref{e:coin_prob} or $\widetilde{\mathcal{R}}_{\rm coin}$ in Eq.~\eqref{e:coin_prob'}, respectively, can be interpreted as a normalized or unnormalized reference probability distribution. However, Eq.~\eqref{e:coin_prob''} can still be given a probabilistic interpretation.

Indeed, by multiplying Eq.~\eqref{e:coin_prob''} by a large natural number $M$ we can interpret each term as a generalized ensemble made up of samples and anti-samples. The vector $M\mcP_{\rm coin}$ would represent the original ensemble of biased coins. The vector $M\widetilde{\mathcal{N}}_{\rm coin}^{(1)}$ would represent an ensemble containing on average $2 M$ samples in state heads and $M p$ anti-samples in state tails. Similarly, the vector $M\widetilde{\mathcal{N}}_{\rm coin}^{(2)}$ would represent an ensemble with, on average, $M(1 + p)$ anti-samples in state heads and $2 M p$ samples in state tails. The sum of these two vectors in Eq.~\eqref{e:coin_prob''} anihilates pairs of samples and anti-samples in the same state, leading to the original ensemble $M\mcP_{\rm coin}$. 

To apply these ideas iteratively, sample by sample, we can take the absolute value of all entries in each expression $\widetilde{\mathcal{N}}^{(1)}$ and $\widetilde{\mathcal{N}}^{(2)}$~\cite{burgin2010interpretations,abramsky2014operational}. The sign tells us whether the corresponding entry corresponds to a sample or an anti-sample.  In this way, Eqs.~\eqref{e:Ncoin1} and \eqref{e:Ncoin2} become
\BE
\left|\widetilde{\mathcal{N}}_{\rm coin}^{(1)}\right| &=& (1 + 2 f_1) \left[\frac{1}{2+p} \begin{pmatrix} 2 \\ p\end{pmatrix}\right],\label{e:|Ncoin1|}\\
\left|\widetilde{\mathcal{N}}_{\rm coin}^{(2)}\right| &=&  (1 + 3 f_2) \left[\frac{1}{1+3p}\begin{pmatrix} 1 + p \\ 2p \end{pmatrix}\right],\label{e:|Ncoin2|}
\EE
where $0\leq f_1 = (1+p)/2\leq 1$ and $0\leq f_2 = p\leq 1$. 

To generate a sample or anti-sample from $\widetilde{\mathcal{N}}_{\rm coin}^{(1)}$, we first generate a sample from the probability vector inside square brackets in Eq.~\eqref{e:|Ncoin1|}, i.e., $\left|\widetilde{\mathcal{N}}_{\rm coin}^{(1)}\right|/(1+2f_1)$. If we get heads or tails we save the sample generated as a sample or anti-sample, respectively. Furthermore, with probability $f_1$ we generate two replicas of the sample or anti-sample saved. Were we to repeat this process $M$ times we would get an ensemble with, on average, $M (1+2f_1) = M(2+p)$ generalized samples, i.e. an average of $2 M$ samples in state heads and $Mp$ anti-samples in state tails. This is precisely the ensemble described by $M \widetilde{\mathcal{N}}_{\rm coin}^{(1)}$. 

However, this is not our aim as we still need to take into account vector $\widetilde{\mathcal{N}}_{\rm coin}^{(2)}$. Instead we now generate a sample from the probability vector inside square brackets in Eq.~\eqref{e:|Ncoin2|}, i.e., $\left|\widetilde{\mathcal{N}}_{\rm coin}^{(2)}\right|/(1+3 f_2)$. If we get tails or heads we save the sample generated as a sample or an anti-sample, respectively. Furthermore, with probability $f_2$ we generate three replicas of the sample or anti-sample saved. Were we to repeat this process alone $M$ times, we would get an ensemble with, on average, $M (1+3 f_2) = M(1+3p)$ generalized samples, i.e. an average of $2 M p$ samples in state tails and $(1+p)M$ anti-samples in state heads. The sum in Eq.~\eqref{e:coin_prob''} implies that, taking into account the two processes above, samples and anti-samples in the same state anihilate each other. 

Now, any signed expression, i.e., vector or function, can be written as the substraction of two non-negative expressions. So, terms like $\widetilde{\mathcal{N}}_{\rm coin}^{(1)}$ and $\widetilde{\mathcal{N}}_{\rm coin}^{(2)}$ in Eq.~\eqref{e:coin_prob''} could be interpreted as (un)normalized probability distributions relative to other (un)normalized reference probability distributions.

In principle, we can extend these ideas to general probabilities $\mcP$ and multiple expressions $\widetilde{\mathcal{N}}^{(\alpha)}$, where $\alpha$ is a suitable index. A common example, pointed out by Feynman~\cite{feynman1987negative} (see Eqs.~(4)-(6) therein), is the solution, $p(x,t)$, to the one-dimensional diffusion equation with absorbing boundary conditions. Indeed, if $p(x,t) = 0$ at $x=0$ and $x=\pi$ the solution to the diffusion equation can be written as
\be\label{e:diff}
p(x,t) = \sum_{k=1}^\infty P_k \sin(k x) e^{-k^2 t},
\ee
where $P_k$ are suitable coefficients and we are assuming the diffusion coefficient is $D=1$. Here we can take 
\be
\widetilde{\mathcal{N}}^{(k)}(x) = P_k\sin (kx) e^{-k^2 t} .
\ee
The expressions $\widetilde{\mathcal{N}}^{(k)}$ are interesting in that they have a relatively simple dynamics, i.e., their spatial shape remains constant except for their amplitude, which decay exponentially fast. By associating the negative values of $\widetilde{\mathcal{N}}^{(k)}(x)$ with anti-samples and associating its lack of normalization with replication or removal of samples, we can in principle give a probabilistic interpretation to this quantity along the lines discussed here. 

There may be situations where it is convenient to work with approximations to a probability distribution. For instance, we could truncate the series in Eq.~\eqref{e:diff} or neglect $\mcR_{\rm coin}$ in Eq.~\eqref{e:coin_prob}. As suggested by Eq.~\eqref{e:coin_prob}, this situation can lead to signed probability distributions like $\mathcal{N}_{\rm coin} = \mcP_{\rm coin} - \mathcal{R}_{\rm coin}$. A situation of this kind arises in quantum mechanics, for instance, in the so-called two-level atom approximation. Indeed, a non-relativistic atom interacting with a radiation field can be described in terms of a non-negative transition kernel~\cite{realpe2017modeling,realpe2018cognitive} associated to a so-called stoquastic Hamiltonian. Such an infinite-dimensional kernel can be written in terms of the countably infinite eigenfunctions of the associated Hamiltonian operator. However, truncating the Hamiltonian operator to only two eigenfunctions, i.e., two energy levels, leads to an effective two-dimensional transition kernel with some negative entries, which is associated to a so-called non-stoquastic Hamiltonian~\cite{realpe2017modeling,realpe2018cognitive} (see, e.g., Appendix~E in Ref.~\cite{realpe2017modeling}). 

As suggested by the memory-enhanced stochastic algorithm introduced in Ref.~\cite{realpe2019quantum}, the use of negative probabilities can have substantial practical value. This algorithm is based on the case where the signed probabilities can be interpreted as relative to a properly normalized probabilistic expression, i.e., the stationary state of a Markov chain. This is the case associated to Eq.~\eqref{e:coin_prob}. It is not clear to us what could be the value of the more general cases illustrated with Eqs.~\eqref{e:coin_prob'} and \eqref{e:coin_prob''}.  Computing normalization constants is one of the bottlenecks in sampling, e.g., in unsupervised machine learning applications~\cite{Goodfellow-et-al-2016}. We wonder if allowing for signed and unnormalized probability distributions could lead to more efficient (even if approximated) sampling algorithms in those cases.


\end{document}